\documentclass[english]{IEEEtran}
\usepackage{lmodern}
\usepackage{lmodern}

\usepackage[T1]{fontenc}
\usepackage[latin9]{inputenc}
\usepackage{babel}
\usepackage{float}
\usepackage{amsmath}
\usepackage{amsthm}
\usepackage{amssymb}
\usepackage{graphicx}
\usepackage[unicode=true,
 bookmarks=true,bookmarksnumbered=true,bookmarksopen=true,bookmarksopenlevel=1,
 breaklinks=false,pdfborder={0 0 0},pdfborderstyle={},backref=false,colorlinks=false]
 {hyperref}
\hypersetup{pdftitle={Your Title},
 pdfauthor={Your Name},
 pdfborderstyle={},pdfpagelayout=OneColumn,pdfnewwindow=true,pdfstartview=XYZ,plainpages=false}

\makeatletter

\floatstyle{ruled}
\newfloat{algorithm}{tbp}{loa}
\providecommand{\algorithmname}{Algorithm}
\floatname{algorithm}{\protect\algorithmname}

\theoremstyle{plain}
\newtheorem{thm}{\protect\theoremname}
\theoremstyle{plain}
\newtheorem{lem}[thm]{\protect\lemmaname}
\theoremstyle{plain}
\newtheorem{prop}[thm]{\protect\propositionname}

\usepackage{babel}
\usepackage{amsthm}
\usepackage{color}

\floatstyle{ruled}
\newfloat{algorithm}{tbp}{loa}
\providecommand{\algorithmname}{Algorithm}
\floatname{algorithm}{\protect\algorithmname}
\theoremstyle{plain}
\theoremstyle{plain}
\theoremstyle{plain}
\usepackage{algorithm}
\usepackage{algorithmic}

\let\rem\@undefined
\theoremstyle{remark}
\newtheorem{rem}{\protect\remarkname}

\let\lem\@undefined
\theoremstyle{lemma}
\newtheorem{lem}{\protect\lemmaname}

\let\prop\@undefined
\theoremstyle{proposition}
\newtheorem{prop}{\protect\propositionname}
\makeatother

\providecommand{\lemmaname}{Lemma}
\providecommand{\propositionname}{Proposition}
\providecommand{\theoremname}{Theorem}

\begin{document}

\title{Analysis and Optimization for Weighted Sum Rate in Energy Harvesting
Cooperative NOMA Systems}

\author{Binh Van Nguyen, Quang-Doanh Vu, and Kiseon Kim \thanks{Binh Van Nguyen and Kiseon Kim are with the School of Electrical Engineering
and Computer Science, Gwangju Institute of Science and Technology,
Republic of Korea. (E-mail : \{binhnguyen, kskim\}@gist.ac.kr).} \thanks{Quang-Doanh Vu is with the Centre for Wireless Communications, University
of Oulu, Finland. (E-mail: doanh.vu@oulu.fi).} \thanks{The authors gratefully acknowledge the support from Electronic Warfare
Research Center at Gwangju Institute of Science and Technology (GIST),
originally funded by Defense Acquisition Program Administration (DAPA)
and Agency for Defense Development (ADD).} }
\maketitle
\begin{abstract}
We consider a cooperative non-orthogonal multiple access system with
radio frequency energy harvesting, in which a user with good channel
harvests energy from its received signal and serves as a decode-and-forward
relay for enhancing the performance of a user with poor channel. We
here aim at maximizing the weighted sum rate of the system by optimizing
the power allocation coefficient used at the source and the power splitting coefficient
used at the user with good channel. By exploiting the specific structure
of the considered problem, we propose a low-complexity one-dimensional
search algorithm which can provide optimal solution to the problem.
As a benchmark comparison, we derive analytic expressions and simple
high signal-to-noise ratio (SNR) approximations of the ergodic rates
achieved at two users and their weighted sum with fixed values of the
power allocation and the power splitting coefficients, from which the scaling
of the weighted sum in the high SNR region is revealed. Finally, we provided
numerical results to demonstrate the validity of the optimized scheme.
\end{abstract}


\section{Introduction}

Non-orthogonal multiple access (NOMA) transmission is emerging as
a promising multiple access technique for the next generation of wireless
networks \cite{Zhang'17}. The cornerstone of NOMA is to exploit the
power domain and channel quality difference among users to achieve
multiple access. An issue rising in a NOMA system is that users with
good channel conditions can significantly strengthen their performance,
while the performance of users with bad channel conditions are relatively
poor \cite{Yang'17}. A possible solution for this problem is combining
cooperative communication with NOMA to generate a cooperative NOMA
(C-NOMA) transmission scheme in which users with good channel conditions
operate as relays to strengthen the transmission reliability for users
suffering from bad channel conditions \cite{Ding'15}-\cite{Men'16}.

Recently, radio frequency energy harvesting (RF-EH) has become an
efficient solution to prolong the lifetime of energy-constraint wireless
communication systems \cite{Fafoutis'15}. The advantage of RF-EH
is from the fact that RF signals carry both information and energy
at the same time, i.e.\ RF-EH allows limited-power nodes to scavenge
energy and process information simultaneously \cite{Wang'16}. There
exist two main RF-EH techniques, namely, time switching (TS) and power
splitting (PS). With TS, a receiver switches between energy harvester
and data decoder. With PS, a receiver separates the RF signals into
two parts (one for EH and the other for decoding) by a PS coefficient.
Here, we mainly focus on PS, since PS is considered to be more general
compared to TS \cite{Shi'14}.

Clearly, RF-EH provides more incentives for user cooperation, thus
it is natural to use RF-EH in C-NOMA systems. Representative examples
for this approach are \cite{Liu'16}-\cite{Do'17} where the systems
with one source and multiple users are considered. These two works
proposed user-pair selection schemes and analyze the performance in
terms of outage probability.

In this paper, we investigate the impact of power allocation and PS
coefficients on the performance of C-NOMA systems. Different from \cite{Liu'16}
and \cite{Do'17}, we focus on weighted sum rate of the systems which
has been still relatively open. It is worth mentioning that weighted
sum rate finds many practical applications since it is helpful for
prioritizing users \cite{Chris'08}. Specifically, our main contributions
are as follows.
\begin{itemize}
\item We consider a C-NOMA with RF-EH system where a source communicates with two users.
We first formulate the problem of weighted sum rate
maximization in which power allocation and PS coefficients are the
design parameters. The problem is non-convex whose optimal solution
can be found by the exhaustive two-dimensional (2D) search. Towards
a more efficient solution, we develop an one-dimensional (1D) search
algorithm by exploiting the specific structure of the problem.
\item For a comparison benchmark, we derive closed-form expressions and
high signal-to-noise ratio (SNR) approximations of the ergodic rates
achieved at the two users and their weighted sum with fixed power
allocation and PS coefficients.
\item We numerically demonstrate that optimized power allocation and PS
coefficient can significantly improves the system performance in terms
of weighted sum rate, i.e.\ $45\%$ enhancement when the average
SNR is $10$ dB and the weight ratio is $5$. On the other hand,
the analysis results reveal that the scaling of the weighted sum rate
is $\frac{w_{1}}{2}\text{log}_{2}\left(\text{SNR}\right)$, where
$w_{1}$ is the priority weight of the user with good channel.
\end{itemize}

\section{System Model}
We consider a wireless communication system consisting of a source,
denoted by $\mathrm{S}$, and two users which are associated with
different channel conditions; we denote the user with good channel
by $\mathrm{U}_{1}$, and the one with bad channel by $\mathrm{U}_{2}$.
All nodes are equipped with a single-antenna and operate in the half-duplex
mode. Let $h_{1}$, $h_{2}$, and $h_{3}$ denote the complex channel
coefficient between $\mathrm{S}$ and $\mathrm{U}_{1}$, $\mathrm{S}$
and $\mathrm{U}_{2}$, $\mathrm{U}_{1}$ and $\mathrm{U}_{2}$, respectively.
All channels are assumed to be independent and identically distributed
Rayleigh block fading. From the assumption about channel quality,
we have $g_{1}>g_{2}$ where $g_{i}=\left|h_{i}\right|^{2}$.

We focus on the transmission from $\mathrm{S}$ to the users. The
transmission protocol includes two phases, each of length $T$ in
time unit. In particular, let $x_{i}$, $i\in\{1,2\}$, be the normalized
complex signal for $\mathrm{U}_{i}$, and $P_{\mathrm{S}}$ be the
transmit power at $\mathrm{S}$. In the first phase, $\mathrm{S}$
generates a superimposed signal given by $x_{\mathrm{S}}=\sqrt{\alpha P_{\mathrm{S}}}x_{1}+\sqrt{\left(1-\alpha\right)P_{\mathrm{S}}}x_{2}$,
where $\alpha$ denotes the power allocation coefficient, and broadcasts
$x_{\mathrm{S}}$ to the users. The received signal at $\mathrm{U}_{i}$
during this phase is
\begin{equation}
y_{i}=h_{i}x_{\mathrm{S}}+n_{i}
\end{equation}
where $n_{i}$ is the additive white Gaussian noise (AWGN) with variance
$N_{0}$.

User $\mathrm{U}_{1}$ uses its received signal for decoding $x_{1}$,
harvesting energy, and decoding $x_{2}$. In particular, $\mathrm{U}_{1}$
divides $y_{1}$ into two parts with a PS coefficient
$\rho\in[0,1]$. The first part given by $y_{1}^{\mathrm{eh}}=\sqrt{\rho}y_{1}$
is for harvesting energy, and the second part given by $y_{1}^{\mathrm{ip}}=\sqrt{1-\rho}y_{1}$
is for decoding information. Consequently, the energy harvested at
$\mathrm{U}_{1}$ is \cite{Do'17}
\begin{equation}
E_{1}=T\eta\rho P_{\mathrm{S}}g_{1}
\end{equation}
where $\eta$ denotes the energy conversion efficiency. $\mathrm{U}_{1}$
decodes $x_{2}$ based on $y_{1}^{\mathrm{ip}}$, then applies successive
interference cancellation (SIC) before decoding $x_{1}$. Therefore,
the signal-to-interference-plus-noise ratios (SINRs) for decoding
$x_{2}$ and $x_{1}$ at $\mathrm{U}_{1}$ are

\begin{align}
\gamma_{1}^{x_{2}}(\alpha,\rho) & =\frac{\left(1-\rho\right)\left(1-\alpha\right)P_{\mathrm{S}}g_{1}}{\left(1-\rho\right)\alpha P_{\mathrm{S}}g_{1}+\left(1-\rho\right)N_{0}+\mu N_{0}},\\
\gamma_{1}^{x_{1}}(\alpha,\rho) & =\frac{\left(1-\rho\right)\alpha P_{\mathrm{S}}g_{1}}{\left(1-\rho\right)N_{0}+\mu N_{0}}
\end{align}
respectively. Here, the last term in the denominator of $\gamma_{1}^{x_{2}}(\alpha,\rho)$
and $\gamma_{1}^{x_{1}}(\alpha,\rho)$ are due to the conversion noise
which is assumed to be AWGN with variance $\mu N_{0}$ \cite{Nguyen'18}.

In the second phase, $\mathrm{U}_{1}$ uses the harvested energy $E_{1}$
to transmit $x_{2}$ to $\mathrm{U}_{2}$. The signal received at
$\mathrm{U}_{2}$ during this phase is
\begin{equation}
\tilde{y}_{2}=\sqrt{\rho\eta P_{\mathrm{S}}g_{1}}h_{3}x_{2}+n_{2}.
\end{equation}
We suppose that the maximal ratio combining (MRC) receiver is used
at $\mathrm{U}_{2}$ \cite{Yue'17}. Then the SINR for decoding $x_{2}$
at $\mathrm{U}_{2}$ is
\begin{align}
\gamma_{2}^{\mathrm{MRC}}(\alpha,\rho)=\frac{\left(1-\alpha\right)P_{\mathrm{S}}g_{2}}{\alpha P_{\mathrm{S}}g_{2}+N_{0}+\mu N_{0}}+\frac{\rho\eta P_{\mathrm{S}}g_{1}g_{3}}{N_{0}+\mu N_{0}}.
\end{align}
In summary, the instantaneous achieved rate at $\mathrm{U}_{1}$ and
$\mathrm{U}_{2}$ are $C_{1}\left(\alpha,\rho\right)=\frac{1}{2}\log_{2}\left(1+\gamma_{1}^{x_{1}}(\alpha,\rho)\right)$
and $C_{2}\left(\alpha,\rho\right)=\frac{1}{2}\log_{2}\left(1+\min\left\{ \gamma_{1}^{x_{2}}(\alpha,\rho),\gamma_{2}^{\mathrm{MRC}}(\alpha,\rho)\right\} \right)$,
respectively.

\section{Weighted Sum Rate Optimization}

Our aim is to maximize the weighted sum rate of the system. Particularly,
the optimization problem is formulated as \begin{subequations} \label{prob:wsc_max}
\begin{align}
\mathop{\textrm{maximize}}\limits _{\alpha,\rho} & \ w_{1}C_{1}\left(\alpha,\rho\right)+w_{2}C_{2}\left(\alpha,\rho\right)\label{eq:obj}\\
\text{subject to} & \ 0<\alpha<1,0\le\rho\le1,
\end{align}
\end{subequations}where $w_{1}>0$ and $w_{2}>0$ are the priority
weights. Here we focus on the case $w_{2}>w_{1}$.\footnote{The optimal solution for the case $w_{2}\leq w_{1}$ is trivial, i.e.
it is not difficult to justify that the optimal solution for this
case is $(\alpha=1,\rho=0)$.} A practical example for the considered scenario is that in cellular
network, the user at cell-edge suffering bad channel conditions for
a long time will be assigned a larger weight compared to the one in
near base station area for fairness and/or stability \cite{Chris'08}.

Objective function \eqref{eq:obj} is non-convex with respect to the
related variables. For achieving an optimal solution, a exhaustive
2D search procedure (over $\alpha$ and $\rho$) can be used. Clearly,
doing this is highly complex and inefficient. In the following, by
looking inside the problem, we develop a low-complexity 1D search
algorithm which solves \eqref{prob:wsc_max} optimally.

We start with an useful result stated as follows.
\begin{lem}
\label{lem:opt}Let $(\alpha^{\ast},\rho^{\ast})$ be an optimal of
\eqref{prob:wsc_max}, then
\begin{equation}
C_{2}\left(\alpha^{*},\rho^{*}\right)=\frac{1}{2}\log_{2}\left(1+\gamma_{2}^{\mathrm{MRC}}\left(\alpha^{*},\rho^{*}\right)\right).
\end{equation}
\end{lem}
\begin{IEEEproof}
The lemma can be proved by contradiction. Specifically, suppose that
there exists an optimal point $(\alpha^{\ast},\rho^{\ast})$ such
that
\begin{align}
\log_{2}\left(1+\gamma_{1}^{x_{2}}\left(\alpha^{*},\rho^{*}\right)\right) & <\log_{2}\left(1+\gamma_{2}^{\mathrm{MRC}}\left(\alpha^{*},\rho^{*}\right)\right).
\end{align}
Since $\gamma_{2}^{\mathrm{MRC}}\left(\alpha,\rho\right)$ and $\gamma_{1}^{x_{2}}\left(\alpha,\rho\right)$
are increasing and decreasing functions of $\rho$, we can always
find $\triangle\rho>0$ such that $C_{2}(\alpha^{\ast},\rho^{\ast}-\triangle\rho)>C_{2}(\alpha^{\ast},\rho^{\ast})$.
Moreover, $C_{1}(\alpha^{\ast},\rho^{\ast}-\triangle\rho)>C_{1}(\alpha^{\ast},\rho^{\ast})$
because $\gamma_{1}^{x_{1}}\left(\alpha,\rho\right)$ is a decreasing
function of $\rho$. Consequently, we have $C_{1}(\alpha^{\ast},\rho^{\ast}-\triangle\rho)+C_{2}(\alpha^{\ast},\rho^{\ast}-\triangle\rho)>C_{1}(\alpha^{\ast},\rho^{\ast})+C_{2}(\alpha^{\ast},\rho^{\ast})$,
which contradicts the assumption that $(\alpha^{\ast},\rho^{\ast})$
is an optimal. This completes the proof.
\end{IEEEproof}
From Lemma \ref{lem:opt} and the monotonicity of the logarithmic
function, we can rewrite \eqref{prob:wsc_max} as \begin{subequations}\label{prob:srmax-1}
\begin{align}
\underset{\alpha,\rho}{\textrm{maximize}} & \ f(\alpha,\rho)\label{eq:obj-1}\\
\text{subject to} & \ \gamma_{1}^{x_{2}}\left(\alpha,\rho\right)\geq\gamma_{2}^{\mathrm{MRC}}\left(\alpha,\rho\right)\label{eq:rateconstr}\\
 & \ 0<\alpha<1,0\leq\rho<1
\end{align}
\end{subequations}where $f(\alpha,\rho)\triangleq\left(1+\gamma_{1}^{x_{1}}\left(\alpha,\rho\right)\right)\left(1+\gamma_{2}^{\mathrm{MRC}}\left(\alpha,\rho\right)\right)^{\tilde{w}_{2}}$,
and $\tilde{w}_{2}=w_{2}/w_{1}$. As a further step, we equivalently
rewrite \eqref{prob:srmax-1} as\begin{subequations}\label{prob:wsc-max-2}
\begin{align}
\underset{\alpha,\rho}{\textrm{maximize}} & \ f(\alpha,\rho)\label{eq:obj-1-1}\\
\text{subject to} & \ 0<\alpha<1,0\leq\rho\leq\tilde{\rho}(\alpha),
\end{align}
\end{subequations}where $\tilde{\rho}\left(\alpha\right)=\frac{b-\sqrt{b^{2}-4ac}}{2a}$,
$\bar{\gamma}=P_{\mathrm{S}}/N_{0}$, $a=\frac{\eta\bar{\gamma}g_{1}g_{3}\left(\alpha\bar{\gamma}g_{1}+1\right)}{1+\mu}$,
$b=\frac{\eta\bar{\gamma}g_{1}g_{3}\left(\alpha\bar{\gamma}g_{1}+\mu+1\right)}{1+\mu}-\frac{\left(1-\alpha\right)\bar{\gamma}g_{2}\left(\alpha\bar{\gamma}g_{1}+1\right)}{\alpha\bar{\gamma}g_{2}+\mu+1}+\left(1-\alpha\right)\bar{\gamma}g_{1}$,
and $c=\left(1-\alpha\right)\bar{\gamma}g_{1}-\frac{\left(1-\alpha\right)\bar{\gamma}g_{2}\left(\alpha\bar{\gamma}g_{1}+1+\mu\right)}{\alpha\bar{\gamma}g_{2}+\mu+1}$.
The equivalence can be proved as follows. We first note that the left
hand-side (LHS) of \eqref{eq:rateconstr} monotonically increases
while the right hand-side (RHS) of \eqref{eq:rateconstr} monotonically
decreases with $\rho$. In addition, when $\rho=0$, the RHS is larger
than the LHS due to the assumption $g_{1}>g_{2}$. Moreover, the $\text{RHS}\rightarrow0$
when $\rho\rightarrow1$. Thus, given $\alpha\in(0,1)$, there exists
an unique $\tilde{\rho}(\alpha)\in(0,1)$ such that \eqref{eq:rateconstr}
is satisfied if and only if $\rho\in[0,\tilde{\rho}(\alpha))$. It
is noting that \eqref{eq:rateconstr} can be written as $a\rho^{2}-b\rho+c\geq0$,
from which we yield $\tilde{\rho}(\alpha)$.

We now focus on objective function \eqref{eq:obj-1}. For a given
$\alpha$, \eqref{eq:obj-1} reduces to a function of $\rho$ given
as
\begin{equation}
f_{\alpha}\left(\rho\right)\triangleq\frac{\left(d-e\rho\right)}{t-\rho}\left(p+q\rho\right)^{\tilde{w}_{2}}\label{eq:obj-given-alpha}
\end{equation}
where $d=1+\mu+\alpha\bar{\gamma}g_{1}$, $e=1+\alpha\bar{\gamma}g_{1}$,
$t=1+\mu$, $p=1+\frac{\left(1-\alpha\right)\bar{\gamma}g_{2}}{\alpha\bar{\gamma}g_{2}+1+\mu}$,
$q=\frac{\eta\bar{\gamma}g_{1}g_{3}}{1+\mu}$. We also introduce a
function of $\alpha$ given as
\begin{equation}
\theta\left(\alpha\right)=\beta^{2}-q\tilde{w}_{2}e\left(dp-ept+q\tilde{w}_{2}td\right)\label{eq:theta-alpha}
\end{equation}
where $\beta\left(\alpha\right)=0.5qd\left(\tilde{w}_{2}-1\right)+0.5qet\left(\tilde{w}_{2}+1\right)$.
We have an useful property of $f_{\alpha}(\rho)$ stated as follows.
\begin{prop}
If $\theta(\alpha)>0$ and $\bar{\rho}(\alpha)=\frac{\beta-\sqrt{\theta(\alpha)}}{q\tilde{w}_{2}N_{0}v}\in(0,1)$,
when $\rho$ increases, $f_{\alpha}(\rho)$ increases until reaches
a maximum at $\bar{\rho}(\alpha)$ then decreases. If $\theta(\alpha)>0$
and $\bar{\rho}(\alpha)\leq0$, $f_{\alpha}(\rho)$ is decreasing
over $\rho\in(0,1)$. Otherwise, $f_{\alpha}(\rho)$ is increasing
over $\rho\in(0,1)$.
\end{prop}
The proof of the proposition can be easily obtained via the gradient
of $f_{\alpha}(\rho)$ given as
\begin{align}
\frac{\partial f_{\alpha}\left(\rho\right)}{\partial\rho}=\frac{\left[\left(d-et\right)\left(p+q\rho\right)+q\tilde{w}_{2}\left(t-\rho\right)\left(d-e\rho\right)\right]}{\left(p+q\rho\right)^{1-\tilde{w}_{2}}\left(t-\rho\right)^{2}}.
\end{align}
The algebraic steps are skipped for the sake of brevity.

The property allows us to find the optimal value of $\rho$ when the
optimal value $\alpha^{\ast}$ is given as follows. $\rho^{\ast}=\bar{\rho}(\alpha^{\ast})$
if $\theta(\alpha^{\ast})>0$ and $0<\bar{\rho}(\alpha^{\ast})<\tilde{\rho}(\alpha^{\ast})$.
If $\theta(\alpha^{\ast})>0$ and $\bar{\rho}(\alpha^{\ast})<0$,
$\rho^{\ast}=0$. Otherwise $\rho^{\ast}=\tilde{\rho}(\alpha^{\ast})$.
In summary, we outline the proposed 1D search procedure in Algorithm
\ref{alg:onedisearch} which outputs the optimal solution of \eqref{prob:wsc_max}.
\begin{algorithm}[t]
\caption{The 1D search for solving \eqref{prob:wsc_max} optimally.}
\label{alg:onedisearch} \begin{algorithmic}[1]

\STATE For each $\alpha\in(0,1)$, calculate $\tilde{\rho}(\alpha)$
and $\theta(\alpha)$. \IF{$\theta(\alpha)>0$} \STATE Calculate
$\bar{\rho}(\alpha)=\frac{\beta\left(\alpha\right)-\sqrt{\theta(\alpha)}}{q\tilde{w}_{2}e}$
\STATE \textbf{ if} $\bar{\rho}(\alpha)\in(0,\tilde{\rho}(\alpha))$
\textbf{then} $\rho^{\ast}(\alpha)=\bar{\rho}(\alpha)$, \\
 \STATE \textbf{elseif} $\bar{\rho}(\alpha)\leq0$, \textbf{then}
$\rho^{\ast}(\alpha)=0$, \\
 \STATE \textbf{else} $\rho^{\ast}=\tilde{\rho}(\alpha^{\ast})$,
\textbf{end if}. \ELSE \STATE$\rho^{\ast}=\tilde{\rho}(\alpha^{\ast})$
\ENDIF \STATE Output: $\left(\alpha^{\ast},\rho^{\ast}\right)=\underset{(\alpha,\rho^{\ast}(\alpha))}{\arg\max}\ f(\alpha,\rho^{\ast}(\alpha))$
\end{algorithmic}
\end{algorithm}

\section{Ergodic Rate Analysis}

In this section, we derive the ergodic (and their corresponding weighted
sum) rates achieved at the users with fixed values of $\alpha$ and
$\rho$, which can be used as a benchmark in evaluating Algorithm
1.

\subsection{Ergodic Rate of $\mathrm{U}_{1}$}

Let us first derive the ergodic rate of the $\mathrm{U}_{1}$, which
can be expressed as follows \cite{Yue'17}
\begin{align}
C_{1}^{\mathrm{e}}=\frac{1}{2\ln\left(2\right)}\int_{0}^{\infty}\frac{1-F_{X}\left(x\right)}{1+x}dx,
\end{align}
where $X=\frac{\left(1-\rho\right)\alpha\bar{\gamma}g_{1}}{1-\rho+\mu}$,
and $F_{X}\left(x\right)$ denotes the cumulative distributed function
(CDF) of $X$ which is given by
\begin{align}
F_{X}\left(x\right)=1-\exp\left(-\frac{\left(1-\rho+\mu\right)x}{\left(1-\rho\right)\alpha\bar{\gamma}\delta_{1}^{2}}\right),
\end{align}
where $\delta_{i}^{2}$ is the power of the channel $h_{i}$. Plugging
(16) into (15) gives
\begin{align}
C_{1}^{\mathrm{e}}=\frac{1}{2\ln\left(2\right)}\exp\left(\frac{1-\rho+\mu}{\left(1-\rho\right)\alpha\bar{\gamma}\delta_{1}^{2}}\right)\Gamma\left(0,\frac{1-\rho+\mu}{\left(1-\rho\right)\alpha\bar{\gamma}\delta_{1}^{2}}\right),
\end{align}
where $\Gamma\left(x,y\right)$ is the incomplete upper Gamma function.

\subsection{Ergodic Rate of $\mathrm{U}_{2}$}

Similar to (15), we have
\begin{align}
C_{2}^{\mathrm{e}}=\frac{1}{2\ln\left(2\right)}\int_{0}^{\infty}\frac{1-F_{Z}\left(z\right)}{1+z}dz,
\end{align}
where $Z=\min\left\{ \gamma_{1}^{x_{2}}(\alpha,\rho),\gamma_{2}^{\mathrm{MRC}}(\alpha,\rho)\right\} =\min\left\{ Y,W\right\} $
and $F_{Z}\left(z\right)$ can be approximated as
\begin{align}
F_{Z}\left(z\right)\simeq1-\Pr\left[Y>z\right]\Pr\left[W>z\right],
\end{align}
where the correlation between $Y$ and $W$ is ignored. It can be
readily verified that the correlation between $Y$ and $W$ vanishes
in the high SNR region implying that the approximation is tight when
the average SNR goes large. The probability term $\Pr\left[Y>z\right]$
is first derived as {\small{}
\begin{align}
\Pr\left[{Y>z}\right]=\left\{ \begin{array}{l}
\hspace{-2mm}0,\;\;\text{if}\;z\ge\frac{{1-\alpha}}{\alpha},\\
\hspace{-2mm}\exp\left({-\frac{{\left({1-\rho+\mu}\right)z}}{{\bar{\gamma}\delta_{1}^{2}\left({1-\rho}\right)\left({1-\alpha-\alpha z}\right)}}}\right),\;\text{if}\;z<\frac{{1-\alpha}}{\alpha}.
\end{array}\right.
\end{align}
}Secondly, $\Pr\left[{W>z}\right]$ can be expressed as follows
\begin{align}
\Pr\left[{W>z}\right]=1-\int_{0}^{z}{{F_{{W_{1}}}}\left({z-y}\right){f_{{W_{2}}}}\left(y\right)dy},\label{eq:equationU2}
\end{align}
where ${W_{1}}=\frac{{\left({1-\alpha}\right)\bar{\gamma}{g_{2}}}}{{\alpha\bar{\gamma}{g_{2}}+1+\mu}}$,
${W_{2}}=\frac{{\rho\eta\bar{\gamma}{g_{1}}{g_{3}}}}{{1+\mu}}$, and
{\small{}
\begin{align}
 & {F_{{W_{1}}}}\left(z\right)=\left\{ \begin{array}{l}
1,\;\;\text{if}\;\;z\ge\frac{{1-\alpha}}{\alpha},\\
1-\exp\left({-\frac{{\left({1+\mu}\right)z}}{{\bar{\gamma}\delta_{2}^{2}\left({1-\alpha-\alpha z}\right)}}}\right),\;\text{if}\;z<\frac{{1-\alpha}}{\alpha},
\end{array}\right.\\
 & {f_{{W_{2}}}}\left(z\right)=2\frac{{\left({1+\mu}\right)}}{{\rho\eta\bar{\gamma}\delta_{1}^{2}\delta_{3}^{2}}}{K_{0}}\left({2\sqrt{\frac{{\left({1+\mu}\right)z}}{{\rho\eta\bar{\gamma}\delta_{1}^{2}\delta_{3}^{2}}}}}\right).
\end{align}
where} ${K_{i}}\left(x\right)$ denotes the modified Bessel function
of the second kind of order $i^{\textrm{th}}$. We note that $(z-y)$
is always less than $\frac{1-\alpha}{\alpha}$ when $z<\frac{1-\alpha}{\alpha}$.
On the other hand, when $z\ge\frac{1-\alpha}{\alpha}$, $z-y\ge\frac{1-\alpha}{\alpha}$
if $y\le z-\frac{1-\alpha}{\alpha}$ and $z-y<\frac{1-\alpha}{\alpha}$
if $z-\frac{1-\alpha}{\alpha}\le y\le z$. Base on this fact, we can
further extend \eqref{eq:equationU2} as follows
\begin{align}
 & \Pr\left[{W>z}\right]=1-{F_{{W_{2}}}}\left(z\right)\nonumber \\
 & +\int_{L\left(z\right)}^{z}{\exp\left({-\frac{{\left({1+\mu}\right)\left({z-y}\right)}}{{\bar{\gamma}\delta_{2}^{2}\left({1-\alpha-\alpha z+\alpha y}\right)}}}\right){f_{{W_{2}}}}\left(y\right)dy},
\end{align}
where $L\left(z\right)=0$ if $z<\left({1-\alpha}\right)/\alpha$,
$L\left(z\right)=z-\left({1-\alpha}\right)/\alpha$ otherwise, and
\begin{align}
{F_{{W_{2}}}}\left(z\right)=1-2\sqrt{\frac{{\left({1+\mu}\right)z}}{{\rho\eta\bar{\gamma}\delta_{1}^{2}\delta_{3}^{2}}}}{K_{1}}\left({2\sqrt{\frac{{\left({1+\mu}\right)z}}{{\rho\eta\bar{\gamma}\delta_{1}^{2}\delta_{3}^{2}}}}}\right).
\end{align}
Plugging (24) and (20) into (19) and (18), we obtain
\begin{align}
C_{2}^{e} & \simeq\int_{0}^{\frac{{1-\alpha}}{\alpha}}{\frac{2}{{1+x}}\exp\left({-\frac{{\left({1-\rho+\mu}\right)z}}{{\bar{\gamma}\delta_{1}^{2}\left({1-\rho}\right)\left({1-\alpha-\alpha z}\right)}}}\right)}\nonumber \\
 & \;\;\;\cdot\sqrt{\frac{{\left({1+\mu}\right)z}}{{\rho\eta\bar{\gamma}\delta_{1}^{2}\delta_{3}^{2}}}}{K_{1}}\left({2\sqrt{\frac{{\left({1+\mu}\right)z}}{{\rho\eta\bar{\gamma}\delta_{1}^{2}\delta_{3}^{2}}}}}\right)dz\nonumber \\
 & \;\;\;+\int_{0}^{\frac{{1-\alpha}}{\alpha}}{\int_{0}^{z}{\exp\left({-\frac{{\left({1-\rho+\mu}\right)z}}{{\bar{\gamma}\delta_{1}^{2}\left({1-\rho}\right)\left({1-\alpha-\alpha z}\right)}}}\right)}}\nonumber \\
 & \;\;\;\cdot\exp\left({-\frac{{\left({1+\mu}\right)\left({z-y}\right)}}{{\bar{\gamma}\delta_{2}^{2}\left({1-\alpha-\alpha z+\alpha y}\right)}}}\right)\nonumber \\
 & \;\;\;\cdot\frac{{2\left({1+\mu}\right)}}{{\rho\eta\bar{\gamma}\delta_{1}^{2}\delta_{3}^{2}\left({1+x}\right)}}{K_{0}}\left({2\sqrt{\frac{{\left({1+\mu}\right)z}}{{\rho\eta\bar{\gamma}\delta_{1}^{2}\delta_{3}^{2}}}}}\right)dydz.
\end{align}
It is worthy noting that (26) can be readily evaluated by using standard
mathematical programs such as Matlab and Mathematica. In addition,
from (17) and (26), we can straightforwardly obtain the system weighted
sum rate, i.e. $C_{\mathrm{sum}}^{\mathrm{e}}=w_{1}C_{1}^{\mathrm{e}}+w_{2}C_{2}^{\mathrm{e}}$,
with fixed value of $\alpha$ and $\rho$.

\subsection{High SNR Analysis}

To gain novel insights from our afore-presented analytic results,
we now investigate the ergodic rates in the high SNR region.
\begin{prop}
\label{prop:In-the-high}In the high SNR region, the ergodic rates
of $\mathrm{U}_{1}$ and $\mathrm{U}_{2}$ can be approximated as
follows
\begin{align}
C_{1}^{\mathrm{e}} & \approx\frac{1}{{2\ln\left(2\right)}}\left[{-\chi-\ln\left({\frac{{1-\rho+\mu}}{{\left({1-\rho}\right)\alpha\delta_{1}^{2}\bar{\gamma}}}}\right)+\frac{{1-\rho+\mu}}{{\left({1-\rho}\right)\alpha\delta_{1}^{2}\bar{\gamma}}}}\right],\\
C_{2}^{\mathrm{e}} & \approx\frac{1}{2}{\log_{2}}\left({1+\frac{{1-\alpha}}{\alpha}}\right),
\end{align}
where $\chi$ denote the Euler constant.
\end{prop}
\begin{IEEEproof}
For $C_{1}^{\mathrm{e}}$, we first note that $\Gamma\left({0,x}\right)=-Ei\left({-x}\right)$,
where $Ei\left({x}\right)$ denotes the exponential integral function.
Then using the the facts that $\exp(x)\xrightarrow{x\rightarrow0}1$
and $Ei\left(x\right)\xrightarrow{x\rightarrow0}\chi+\ln\left(-x\right)+x$,
we can obtain (27). For $C_{2}^{\mathrm{e}}$, let's first recall
its instantaneous expression ${C_{2}}=\frac{1}{2}{\log_{2}}\left({1+\min\left\{ {\gamma_{1}^{{x_{2}}}(\alpha,\rho),\gamma_{2}^{\mathrm{MRC}}(\alpha,\rho)}\right\} }\right)$.
Then, in the high region of $\bar{\gamma}$, we can readily show that
$\gamma_{1}^{{x_{2}}}(\alpha,\rho)\to\frac{{1-\alpha}}{\alpha}<\gamma_{2}^{\mathrm{MRC}}(\alpha,\rho)\to\frac{{1-\alpha}}{\alpha}+\frac{{\rho\eta\bar{\gamma}{g_{1}}{g_{3}}}}{{1+\mu}}$,
from which (28) can be obtained.
\end{IEEEproof}
Proposition \ref{prop:In-the-high} implies that as the average SNR
$\bar{\gamma}$ increases, the ergodic rate of $\mathrm{U}_{1}$ monotonically
increases, however, that of $\mathrm{U}_{2}$ is saturated. This is
reasonable because as $\bar{\gamma}$ increases, the SNR used for
decoding $x_{1}$ at $U_{1}$ also increases, and thus, the ergodic
rate of $\mathrm{U}_{1}$ increases. On the other hand, the actual
SINR used for decoding $x_{2}$ is limited by the minimum of the SINRs
used for decoding $x_{2}$ at $\mathrm{U}_{1}$ and $\mathrm{U}_{2}$.
In addition, when $\bar{\gamma}$ increases, the SINR used for decoding
$x_{2}$ at $\mathrm{U}_{1}$ quickly converges to $\frac{1-\alpha}{\alpha}$
and limits the actual SINR used for decoding $x_{2}$, which makes
the ergodic rate of $\mathrm{U}_{2}$ saturated.

From Proposition \ref{prop:In-the-high}, we have
\begin{align}
 & C_{\mathrm{sum}}^{\mathrm{e}}={w_{1}}C_{1}^{\mathrm{e}}+{w_{2}}C_{2}^{\mathrm{e}}\nonumber \\
 & \approx\frac{{w_{1}}}{2}{\log_{2}}\left({\bar{\gamma}}\right)+\frac{w_{2}}{2}\log_{2}\left(\frac{1}{\alpha}\right)\approx\frac{{w_{1}}}{2}{\log_{2}}\left({\bar{\gamma}}\right),
\end{align}
which reveals that when $\bar{\gamma}\rightarrow\infty$, the scaling
of the system weighted sum rate is $\frac{w_{1}}{2}\log_{2}\left(\bar{\gamma}\right)$.

\section{Numerical Results and Discussions}
In this section, we provide representative simulated and analytical
results to validate our analysis and demonstrate the enhancement of
the system performance achieved by the proposed 1D search algorithm.
The simulation setup follows the system model given in Section II
with $\eta=1$ and $\delta_{1}^{2}=\delta_{2}^{2}=\delta_{3}^{2}=1$.
\begin{figure}[t]
\centering \includegraphics[width=6.5cm]{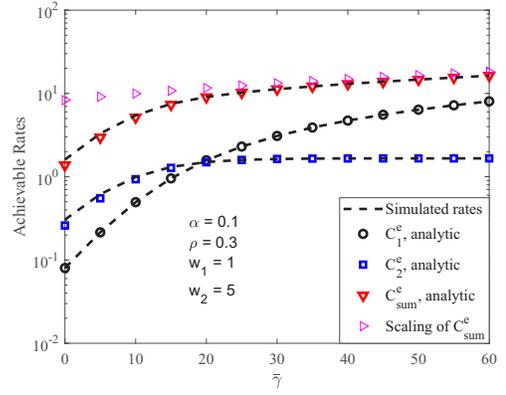} \caption{Achievable rates with fixed values of $\alpha$ and $\rho$.}
\end{figure}

Figure 1 plots the ergodic rates of the considered system with fixed
values of $\alpha$ and $\rho$. The first observation is that the
analytic curve of $C_{1}^{\mathrm{e}}$ follows the corresponding
simulated one excellently, while the analytic curves of $C_{2}^{\mathrm{e}}$
and $C_{\mathrm{sum}}^{\mathrm{e}}$ quickly converge to the corresponding
simulated curves in the medium and high SNR regions. This result implies
that our analyses on the system's ergodic rates are valid. Clearly,
the figure confirms our finding on the scaling of the weighted sum
rate in the high SNR region. The other interesting observation is
that the ergodic rate of $\mathrm{U}_{2}$ is saturated as the average
SNR gets large, revealing that increasing the average SNR (or equally
increasing the transmit power $P_{\mathrm{S}}$) cannot enhance the
performance of the user with poor channel.

Figure 2 plots the system weighted sum rates with optimal and fixed
values of $\alpha$ and $\rho$ as functions of the average SNR. We
take $\tilde{w}_{2}=\{2,5\}$. The figure clearly shows that using
Algorithm 1 remarkably enhances the weighted sum rate performance
of the system. Particularly, at $\bar{\gamma}=10$ dB, optimal values
of $\alpha$ and $\rho$ provides $45\%$ and $29.3\%$ weighted sum
rate enhancements with $\tilde{w}_{2}=5$ and $\tilde{w}_{2}=2$.
Thus, the results strongly suggest that parameter $\alpha$ and $\rho$
should be optimized.

In Fig. 3, we illustrate the average of the optimal values of $\alpha$
and $\rho$ (i.e.\ $\mathbb{E}\left\{ a^{*}\right\} $ and $\mathbb{E}\left\{ \rho^{*}\right\} $,
respectively) versus $\tilde{w}_{2}$. An observation is that as $\tilde{w}_{2}$
increases, $\mathbb{E}\left\{ a^{*}\right\} $ reduces and approaches
zero. This is due to the fact that when $\tilde{w}_{2}$ enlarges,
$\mathrm{U}_{2}$ has a higher priority compared to $\mathrm{U}_{1}$,
and thus, more power should be allocated to the transmission of $x_{2}$.
On the other hand, we can also observe that $\mathbb{E}\left\{ \rho^{*}\right\} $
increases and tends to a certain value. This is because the rate of
$\mathrm{U}_{2}$ provided in Lemma 1 is an increasing function with
$\rho$, and $\rho^{\ast}$ should be small enough so that constraint
(10b) is satisfied.
\begin{figure}[t]
\centering \includegraphics[width=6.5cm]{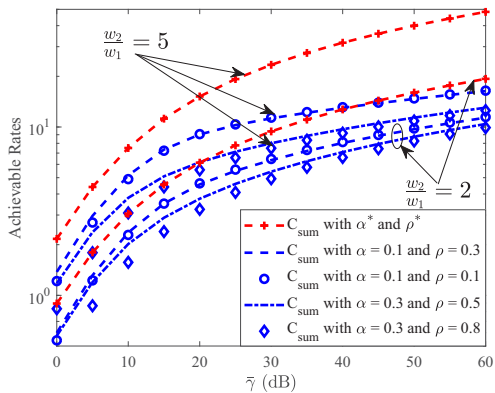} \caption{Weighted sum rate with optimal and fixed values of $\alpha$ and $\rho$.}
\end{figure}

\begin{figure}[t]
\centering \includegraphics[width=6.5cm]{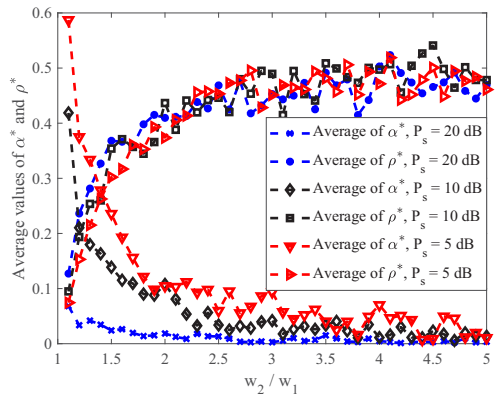} \caption{Average value of $\rho^{\ast}$ and $\alpha^{\ast}$ versus $\tilde{w}_{2}$.}
\end{figure}

\section{Conclusion}
We considered a C-NOMA system with RF-EH including a source and two
users. We first developed a 1D search algorithm to optimally solve
the problem of weighted sum rate maximization respect to power allocation
$\alpha$ and power splitting coefficient $\rho$. Then, we derived
closed-form expressions and high SNR approximations of the ergodic
rates achieved at the two users with fixed values of $\alpha$ and
$\rho$. The numerical results demonstrated that using the optimal
values of $\alpha$ and $\rho$ significantly enlarges the system
weighted sum rate, i.e.\ $45\%$ enhancement when the average SNR
is $10$ dB and the weight ratio is $5$. In addition, from analytic
results, we revealed that the scaling of the weighted sum with fixed
value of $\alpha$ and $\rho$ is $\frac{w_{1}}{2}\log_{2}\left(\bar{\gamma}\right)$.

\end{document}